\documentclass[%
 reprint,
 amsmath,amssymb,
 aps,
 prb,
]{revtex4-1}

\usepackage{graphicx}
\usepackage{dcolumn}
\usepackage{bm}
\usepackage{gensymb} 

\begin{document}

\preprint{APS/123-QED}

\title{Alkali doping of graphene: the crucial role of high temperature annealing
}

\author{A. Khademi}
	\affiliation{Stewart Blusson Quantum Matter Institute, University of British Columbia, Vancouver, BC, V6T1Z4, Canada}
	\affiliation{Department of Physics and Astronomy, University of British Columbia, Vancouver, BC, V6T1Z1, Canada}
\author{E. Sajadi}
	\affiliation{Stewart Blusson Quantum Matter Institute, University of British Columbia, Vancouver, BC, V6T1Z4, Canada}
	\affiliation{Department of Physics and Astronomy, University of British Columbia, Vancouver, BC, V6T1Z1, Canada}
\author{P. Dosanjh}
	\affiliation{Stewart Blusson Quantum Matter Institute, University of British Columbia, Vancouver, BC, V6T1Z4, Canada}
	\affiliation{Department of Physics and Astronomy, University of British Columbia, Vancouver, BC, V6T1Z1, Canada}
    \author{D. Bonn}
	\affiliation{Stewart Blusson Quantum Matter Institute, University of British Columbia, Vancouver, BC, V6T1Z4, Canada}
	\affiliation{Department of Physics and Astronomy, University of British Columbia, Vancouver, BC, V6T1Z1, Canada}
\author{J. A. Folk}
\email{jfolk@physics.ubc.ca}
	\affiliation{Stewart Blusson Quantum Matter Institute, University of British Columbia, Vancouver, BC, V6T1Z4, Canada}
	\affiliation{Department of Physics and Astronomy, University of British Columbia, Vancouver, BC, V6T1Z1, Canada}
\author{A. St{\"o}hr}
	\affiliation{Max Planck Institute for Solid State Research, 70569 Stuttgart, Germany}
\author{S. Forti}
    \affiliation{Max Planck Institute for Solid State Research, 70569 Stuttgart, Germany}
	\affiliation{Present address: Centre for Nanotechnology Innovation IIT@NEST, Piazza San Silvestro 12, 56127 Pisa, Italy}
\author{U. Starke}
	\affiliation{Max Planck Institute for Solid State Research, 70569 Stuttgart, Germany}




\date{\today}

\begin{abstract}
The doping efficiency of lithium deposited at cryogenic temperatures on epitaxial and CVD monolayer graphene has been investigated under ultra-high vacuum conditions. Change of charge carrier density was monitored by gate voltage shift of the Dirac point and by Hall measurements, in low and high doping regimes. It was found that pre-annealing the graphene greatly enhanced the maximum levels of doping that could be achieved: doping saturated at $\Delta n = 2\times 10^{13}$ e$^-$/cm$^2$ without annealing, independent of sample type or previous processing; after a 900 K anneal, the saturated doping rose one order of magnitude to $\Delta n = 2\times 10^{14}$ e$^-$/cm$^2$.
\begin{description}
\item[PACS numbers]
May be entered using the \verb+\pacs{#1}+ command.
\end{description}
\end{abstract}

\pacs{Valid PACS appear here}
\maketitle



Graphene, as an atom-thin surface conductor, offers the unique possibility of tailoring electronic properties by depositing adatoms (or molecules) directly on the carbon lattice.  By proper choice of adatom, it may be possible to open a band gap \cite{Graphane.Science2009, fluorine.PhysRevB.2010, flourine.NanoLett2010}; create local magnetic moments \cite{Flourine.PhysRevLett2012, 3dTransition.PhysRevLett.2013}; enhance spin-orbit coupling to the point the graphene may become a quantum spin Hall insulator \cite{Franz.PhysRevX.1,Franz.PhysRevLett.2012} or experience a quantum anomalous Hall effect \cite{QAHE.PhysRevLett.2012}; or even to induce superconductivity \cite{LiSC.NatPhys2012,SC.Alkali.honeycomb.PhysRevB.2015, Plasmon.SC.PhysRevLett.2007, Chiral.SC.Nat.Phys2012}.
Alkaline adatoms in particular (alkali metals or alkaline earth metals) should dope graphene very efficiently, approaching one extra electron per carbon atom at monolayer coverage. Alkali on graphene are predicted to enhance electron-phonon coupling to the point that superconducting critical temperatures of several Kelvin or more are achieved,\cite{LiSC.NatPhys2012,SC.Alkali.honeycomb.PhysRevB.2015} due both to heavy doping with an associated increase in the  electronic density of states, as well as  changes to the deformation potential and phonon frequency.\cite{LiSC.NatPhys2012,Li.SC.EPL.2014,DFT.Metals.PhysRevB.2008}

The promise of adatom alterations to graphene's electronic structure has been realized in some experiments, whereas several others have reported a surprising absence of adatoms' predicted effects. For example, adatoms have been confirmed to cause charged-impurity scattering in graphene \cite{K.Nat.Phys.2008,K.PhysRevLett.2011,Ca.SolidStateCommunications2012} consistent with theoretical predictions \cite{AdamPNAS2007, Impurity.Scattering.J.Phys.Soc.Jpn.2006, Impurity.Scattering.PhysRevLett.2007, Impurity.Scattering.PhysRevLett.2006, min.cond.PhysRevLett.2007}. At the same time, the theoretically predicted enhancement of spin-orbit interaction on graphene by heavy metal adatoms like indium \cite{Franz.PhysRevX.1,Franz.PhysRevLett.2012,In.Tl.SOC.PhysRevB.2015, In.Tl.SOC.PhysRevLett.2015-2} has not been observed experimentally \cite{In.5K.PhysRevB.2015,In.12K.PhysRevB.2015}.

One of the challenges to moving forward in this area is the difficulty of comparing experiments performed using different forms of graphene, different substrates, different preparation techniques and deposition conditions, even different adatoms, to each other or to theoretical predictions. It is well established, for example, that a given adatom species deposited on single layer graphene may remain on top of the graphene sheet at its interface with vacuum \cite{Fedorov.ARPES.Alkali.Nat.Commun2014, Li.RT.Graphene/CoSi2015}, or may intercalate underneath between the graphene and its substrate \cite{Li.intercalated.monolayer.PhysRevB.2010, Li.intercalation.SiC.PhysRevB.2015, Li.intercalation.H.SiC.Surf.Sci.2012}, depending on the preparation conditions.


Here, we report a transport investigation of what is arguably the simplest of all adatom effects on graphene: charge doping by alkali atoms (Li) deposited under cryogenic UHV conditions. Earlier measurements of the Fermi surface area for Li-on-graphene, using angle-resolved photoemission (ARPES), confirmed that each Li adatom contributed approximately one electron to the graphene carrier density, for both cryogenic and elevated temperature deposition conditions, though the carrier density was observed to saturate to values differing by a factor of nearly 5 in the two experiments\cite{Bart,Fedorov.ARPES.Alkali.Nat.Commun2014}. Transport measurements of dilute potassium and calcium adatoms on graphene, also deposited under cryogenic UHV conditions, observed charged-impurity scattering due to the adatoms; these measurements were consistent with a similar charge transfer $\sim$1e$^{-}$/adatom. However, doping in these experiments was not explored beyond approximately 0.1\% of full coverage \cite{K.Nat.Phys.2008,K.PhysRevLett.2011,Ca.SolidStateCommunications2012}.
%

The primary outcome of our experiment is the observation that alkali doping on graphene, after deposition under cryogenic UHV conditions, in general saturates far below the values reported in Refs.~\onlinecite{Fedorov.ARPES.Alkali.Nat.Commun2014} and \onlinecite{Bart}.  Different types of graphene, prepared with and without resist-based processing, saturate at a $2\times 10^{13}$ e$^-$/cm$^2$ doping level after Li deposition at 4 K. Only when an {\it in situ} annealing step to 900 K has been performed, prior to the cryogenic deposition, is the full doping expected for monolayer alkali coverage recovered.  Apparently, the UHV bake-out process and even subsequent annealing to 500 or 700 K fails to prepare a pristine graphene surface when close adatom-graphene interactions are required. A comparison of annealing protocols sheds light on possible explanations for this effect.



\begin{figure}[t]
  \centering
  \includegraphics{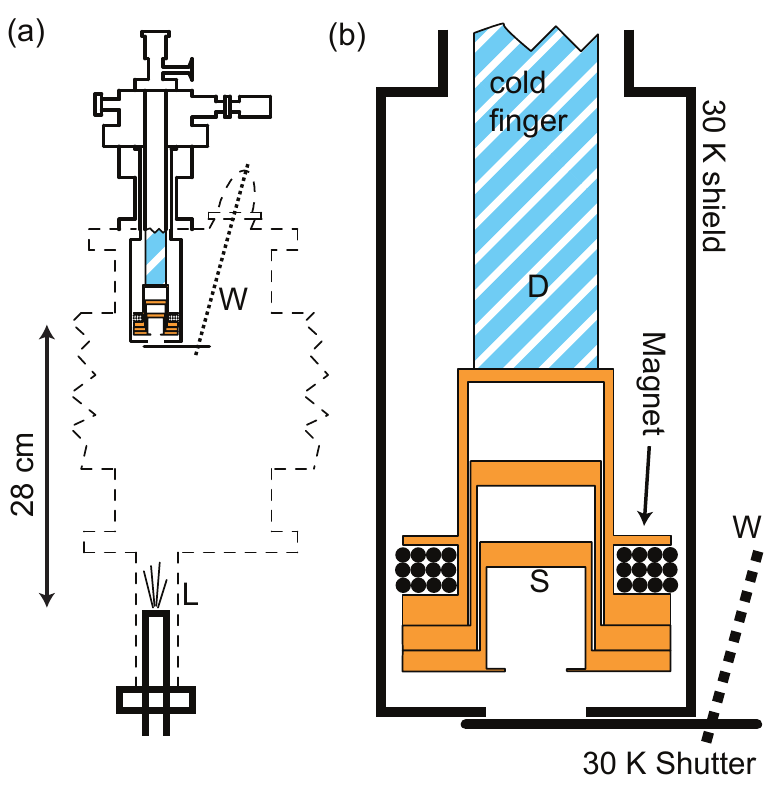}
  \caption{(a) Scaled schematic of experimental set up; dashed line illustrates the walls of UHV chamber, with jagged segments showing connection to the pumps and other parts of chamber. L: Li source, W: wobble stick. (b) a close up view of sample stage in panel (a); S: sample, D: diode.}
  \label{FigSetup1}
\end{figure}

Measurements were performed on seven different graphene devices: five epitaxial graphene samples on SiC (SiC1-5), and 2 samples grown by CVD and transferred on to SiO$_2$/Si chips (CVD1,2). SiC3,4,5 were measured with and without performing annealing tests, whereas SiC1,2 and CVD1,2 were measured directly after bakeout, without a subsequent annealing step.  

SiC1 was a 3$\times$3 mm$^2$ epitaxial monolayer graphene grown on a weakly-doped 6H-SiC(0001) surface \cite{SiC.PhysRevB.84.125449}.
SiC2, SiC3, SiC4 and SiC5, respectively  4$\times$4 mm$^2$, 4$\times$4 mm$^2$, 4$\times$2.5 mm$^2$ and 4$\times$2.5 mm$^2$, were cut from commercially available epitaxial monolayer graphene grown on the semi-insulating 4H-SiC(0001) surface\cite{graphensic}.
Eight contacts were deposited onto the corners and edges of each SiC sample, using shadow evaporation to avoid polymer resist contamination on the as-grown graphene surface. Contacts on SiC1,2,4,5 were 5 nm Cr/90 nm Au; on SiC3 contacts were 100 nm Au.

CVD1 and CVD2 were commercially available monolayer graphene grown by CVD on copper foil: CVD1 was purchased already-transferred\cite{Graphenea} onto a Si/SiO$_{2}$ chip, then etched by oxygen plasma into a Hall bar geometry (29.5 $\mu$m wide with longitudinal $\rm{R_{xx}}$ contacts separated by 73.6 $\mu$m) defined by electron beam lithography (EBL). CVD2 was transferred\cite{BGT} in our laboratory using procedures described in Ref.~\onlinecite{Mathieu.Massicotte.M.Sc.dissertation}, then etched by oxygen plasma into a square ($\sim$700$\times$700 $\mu$m$^2$) defined by EBL. CVD samples were electrically contacted with Ni/Au (5nm/90nm) squares defined by EBL and liftoff. After fabrication, CVD devices were annealed in forming gas (5.72 \% H$_{2}$, balance N$_{2}$) at 350 $\degree$C for 1.5 hours to remove resist residues.

Deposition and transport measurements were carried out on a cryogenic stage in UHV, at pressures below 5$\times$10$^{-10}$ torr after a 3 day bakeout at 100 $\degree$C [Fig.~\ref{FigSetup1}a].
The sample was at the center of a superconducting coil that could be energized to 100 mT. The frame of this magnet plus a cover plate acted as a 4 K shield. The sample stage was screwed to the magnet frame which itself was glued and screwed to a copper stub (cold finger). The cold finger's temperature could be controlled down to 3 K using pumped liquid He, as monitored by a silicon diode (Lakeshore DT-670B-SD). In addition to the 4 K shield, there was a 30 K shield that was closed using a mechanical shutter before and after Li evaporation [Fig.~\ref{FigSetup1}b]. SiC1,2 and CVD1,2 chips were thermally sunk directly to the sample stage. SiC3,4,5 used a customized stage that enabled pre-annealing operations close to 1000 K, while still cooling efficiently for subsequent cryogenic deposition and measurement. In all cases, cryogenic temperatures on the graphene were monitored via the electron-electron contribution to resistivity.

\begin{figure}[t]
  \centering
  \includegraphics{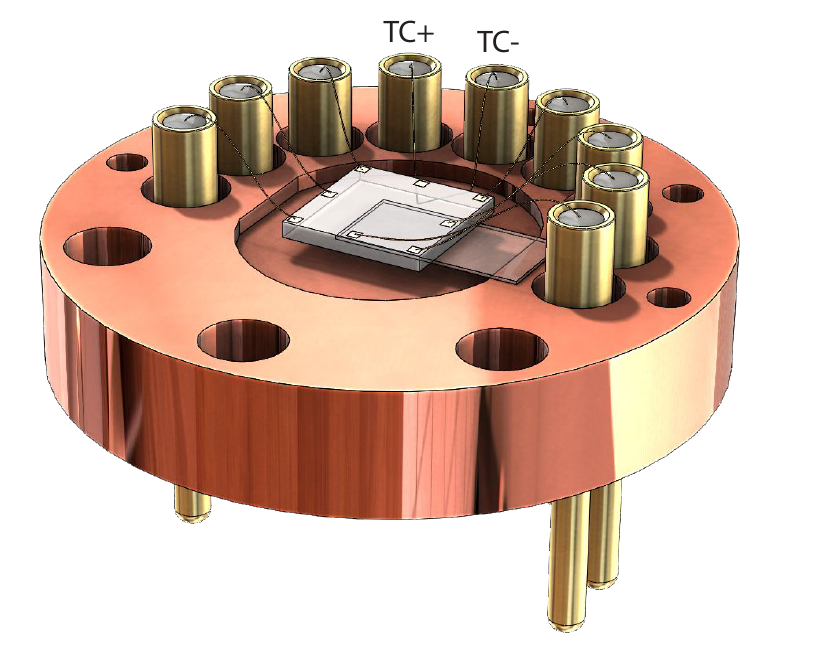}
  \caption{Pre-annealing stage, showing chip glued to end of quartz plate with thermocouple attached to leads TC+ and TC-.}
  \label{Figheaterstage}
\end{figure}

The custom stage for SiC3,4,5 [Fig.~\ref{Figheaterstage}] included a 3$\times$12$\times$0.1 mm single crystal z-cut quartz plate serving as a thermal insulator at high temperature while offering effective thermal coupling at cryogenic temperatures. Samples were glued to one end of the plate, with the opposite end glued to the copper body of the stage.  Annealing temperatures were achieved by driving current through the graphene, from three contacts on one edge of the sample as a source to three on the opposite edge as a drain. Elevated chip temperatures were monitored with a 13 $\mu$m Chromel/Alumel thermocouple between the chip and the quartz plate. 50 mA through the graphene required 25 V, giving 1.25 W dissipation and raising the sample temperature to 900 K while other areas in the sample holder stayed less than 350 K. 


Lithium was evaporated by driving current through an alkali metal dispenser (SAES Getters) located  28 cm (all except CVD1) or 13 cm (CVD1) below the sample stage. Li sources were first degassed for 30 min at 6 A, and 1-2 min in 7-7.5 A, after the bakeout.  In all cases, deposition occurred when the sample temperature was at 4 K. Immediately before evaporation, the source current was raised to the desired level for one minute, then the shutter was opened during deposition.  The sample temperature rose by at most 1 K during evaporation.  Li vapor during evaporation was detected by a residual gas analyser positioned off-axis. The presence of Li on the graphene surface was later confirmed by time-of-flight secondary ion mass spectrometry. 

\begin{figure}[t]
  \centering
  \includegraphics[scale=0.94]{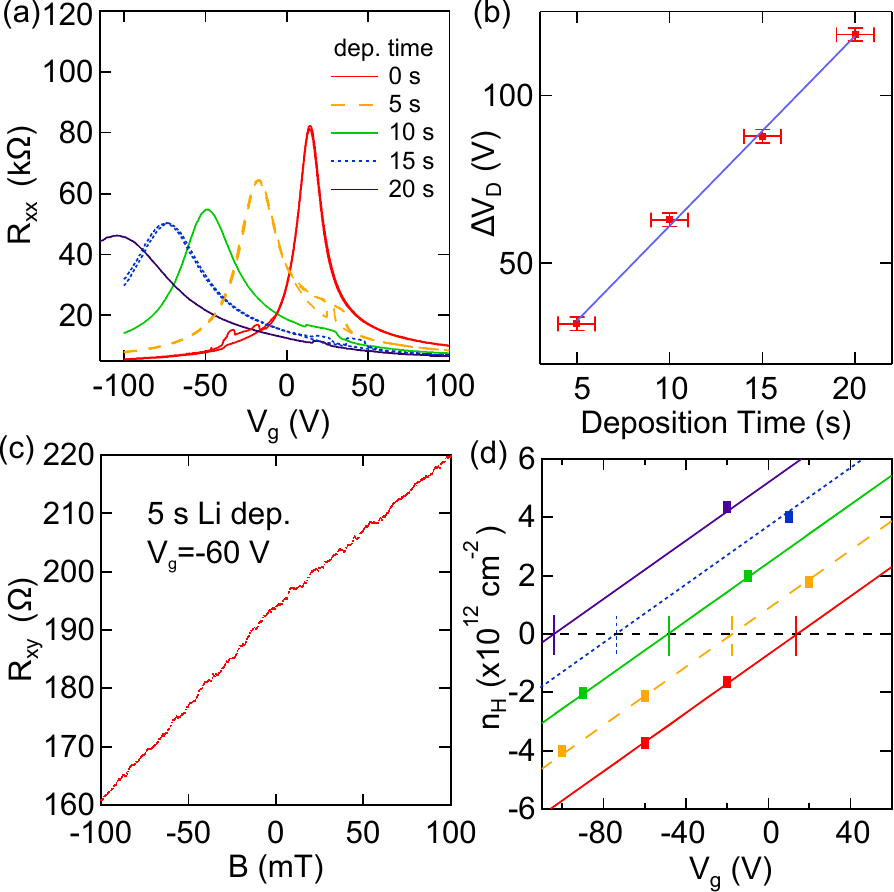}
  \caption{Li deposition at 3 K on CVD1: (a) Dirac peak shift with consecutive Li depositions. Retraced data show repeated scans, confirming no shift of Dirac peak with time after the shutter was closed. Legend indicates total deposition time.  (b) Gate voltage shift of Dirac peak center ($\Delta$V$_{D}$) was linear in deposition time. (c) Example of Hall data. (d) Carrier density, $n_H$, extracted via Hall effect at different gate voltages for each deposition time (marker size indicates error bars). Diagonal lines [legend as in (a)] correspond to backgate capacitance $\alpha$=n$_{H}$/V$_{g}$=5.0$\times$10$^{14}$ m$^{-2}$V$^{-1}$ with zero-crossing (vertical lines) set by Dirac point at each deposition time.}
  \label{FigDepTime}
\end{figure}


Figure 3 illustrates mild electron doping by Li on CVD1, clearly visible as a shift in the Dirac peak (the zero-carrier-density resistance peak) to more negative gate voltages [Fig.~\ref{FigDepTime}a]. Spatial charge inhomogeneity introduced by Li adatoms broadened the Dirac peak with each deposition. For fixed Li source current, the shift in the Dirac peak was linear in deposition time [Fig.~\ref{FigDepTime}b], confirming the constant deposition rate from the getter source over multiple depositions spread over the ten hours required to accumulate a set of data such as that in Fig.~\ref{FigDepTime}.

Dirac peak shift in gate voltage is a useful probe of charge carrier density for mild doping, but is ineffective at higher densities when the Dirac peak moves out of the range of accessible gate voltages, and on the (un-gateable) SiC samples. Induced charge density was also monitored using the Hall effect away from the Dirac point [Fig.~\ref{FigDepTime}c]. Hall and Dirac point measurements were consistent throughout the accessible gate voltage range
[Fig.~\ref{FigDepTime}d].
 
Although the induced carrier density can be determined as in Fig.~\ref{FigDepTime}, the density of adatoms on the surface could not be directly measured.  The ratio between the two is $\eta$, the net charge transferred per adatom.  A range $\eta=0.5-0.7$ has been predicted by DFT calculations, assuming the Li sits on the Hollow site (at the center of a ring of carbon atoms)\cite{DFT.Langmuir2004, DFT.J.Phys.Chem.B2006, DFT.J.Chem.Phys.2005, DFT.PhysRevB.2014}.  Measured values for $\eta$ range from  0.5\cite{Li.RT.Graphene/CoSi2015} to 0.66\cite{Fedorov.ARPES.Alkali.Nat.Commun2014} for ARPES data. 

\begin{figure}[t]
  \centering
  \includegraphics{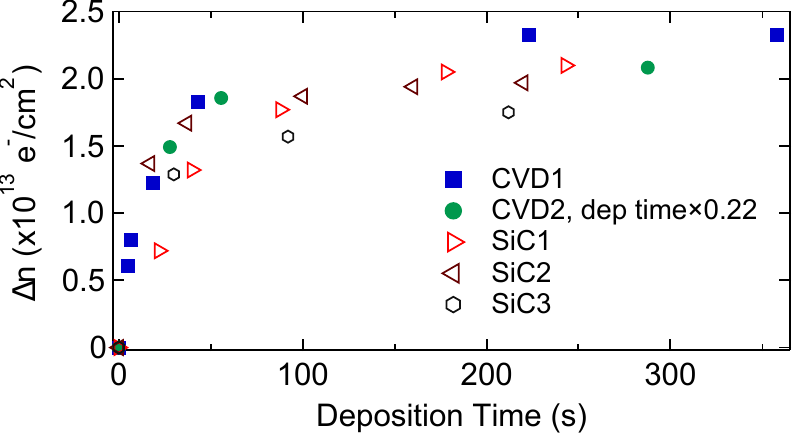}
  \caption{Saturated doping in five samples. Source current was 7.3/7.5 A, except for SiC1 for which it was 7 A. Deposition times reported for CVD2 are multiplied by a geometrical factor (13 cm/28 cm)$^2$ to account for different source-sample distance (see text). n$_{\rm{init}}$ for each deposition was, respectively, 0.4, 0.16, 1.9, 1.7, 1.0$\times10^{13}$ e$^-$/cm$^2$ for CVD1,2, SiC1,2,3.}
  \label{FigSiCDenDepTime}
\end{figure}

For heavier adatom deposition on CVD1 than what is shown in Fig.~\ref{FigDepTime}, the charge carrier density saturated at $2.5\times10^{13}$ e$^-$/cm$^2$ [Fig.~4], an increase of $\Delta \rm{n}_{sat}=2.1\times10^{13}$ e$^-$/cm$^2$ above the initial density $\rm{n}_{init}=0.4\times10^{13}$ e$^-$/cm$^2$ recorded after UHV bakeout but before Li exposure. Nearly identical saturated doping levels, $\Delta \rm{n}_{sat}=2\times10^{13}$ e$^-$/cm$^2$, were observed for CVD2, and SiC1,2,3.  In contrast, an order of magnitude larger $\Delta\rm{n}_{(\sqrt[]{3}\times\sqrt[]{3})R30\degree}=1.9\times10^{14}$ e$^-$/cm$^2$ is expected for the ($\sqrt[]{3}$$\times$$\sqrt[]{3}$)R30$\degree$ arrangement of Li on graphene at the predicted $\eta=0.6$. It is unlikely that the anomalously low doping observed here results from processing, given that CVD1 and 2 followed different processing protocols, and SiC1,2,3 were never exposed to polymer resists.  Also, the saturated doping was apparently not affected by initial carrier density [Fig.~4].

The saturated doping also did not depend on the current passed through the getter, or the getter-to-sample distance.  The graphene-getter distance was 13 cm for CVD2, but 28 cm for CVD1 and SiC1,2,3.  If radiative heating from the source allowed adatoms to move around, for example to dimerize, this effect should have been stronger for CVD2 compared to CVD1.  After scaling deposition time by the geometrical factor (13 cm/28 cm)$^2$ to account for different source-sample distances, even the rate of doping increase was the same [Fig.~\ref{FigSiCDenDepTime}] for CVD1 and CVD2.   The results of each deposition in Fig.~\ref{FigSiCDenDepTime} were reproduced in a second run with the same getter source, confirming that the saturation was not associated with empty Li sources.

\begin{figure}[t]
  \centering
  \includegraphics{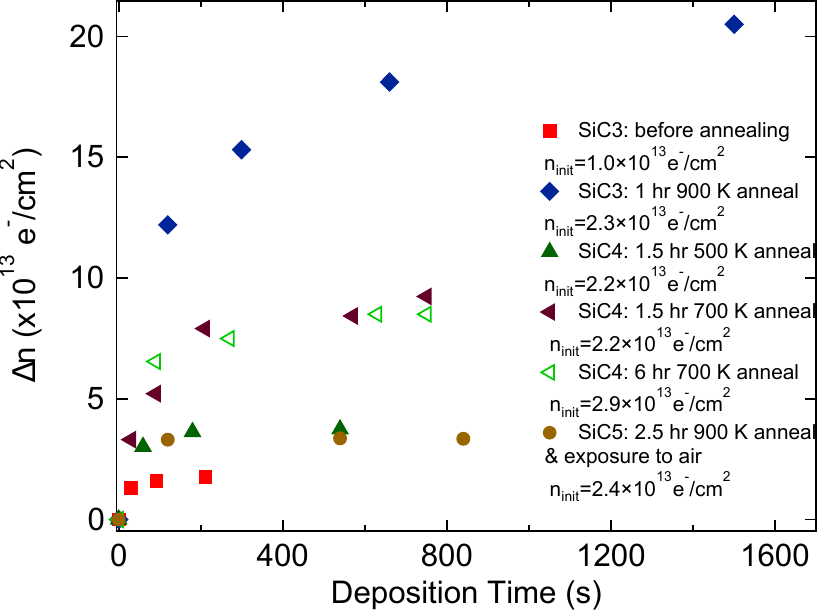}
  \caption{Change of carrier density $\Delta$n versus Li deposition time for SiC3,4,5, before and after annealing. The  Li evaporation source was not changed between subsequent measurements of a given sample; getter current was 7.5 A in all cases.}
  \label{FigSiCDenDepTimeafterannealing}
\end{figure}

The one additional step that did increase $\Delta \rm{n}_{sat}$ was a post-bakeout anneal, prior to cryogenic Li deposition.  Fig.~\ref{FigSiCDenDepTimeafterannealing} illustrates the progressively higher $\Delta \rm{n}_{sat}$ found for pre-deposition annealing temperatures 500 K and above.  In all cases where identical preparations were performed on multiple samples, the same values of $\Delta \rm{n}_{sat}$ were found, validating the comparison of different SiC samples on a single graph.  SiC3 data shows the $\Delta \rm{n}_{sat}=2\times10^{13}$ e$^-$/cm$^2$ baseline before annealing, consistent with Fig.~4, then the order of magnitude larger $\Delta \rm{n}_{sat}=2\times10^{14}$ e$^-$/cm$^2$ that was reached reached when the sample was annealed to 900 K prior to the Li cryogenic deposition.  SiC4 data shows that a 500 K anneal yields $\Delta \rm{n}_{sat}=4\times10^{13}$ e$^-$/cm$^2$, while 700 K yields $\Delta \rm{n}_{sat}=9\times10^{13}$ e$^-$/cm$^2$, and that $\Delta \rm{n}_{sat}$ is independent of annealing time.
We note that the value $\Delta \rm{n}_{sat}=2\times10^{14}$ e$^-$/cm$^2$, recorded in SiC3 after a 900 K anneal, is considerably larger than the $\Delta\rm{n}_{sat}\sim 9\times10^{13}$ e$^-$/cm$^2$ found in Ref.~\onlinecite{Bart} for cryogenic deposition under nearly identical conditions.


How can we understand the saturation of doping at 10\% of the expected level for unprocessed epitaxial graphene after a standard UHV bakeout?  Conversely, what changes are induced on the surface by the 500, 700, and 900 K anneals, that raise the saturated doping levels by an order of magnitude?  In order to address these questions, SiC5 was first annealed at 900 K, then exposed to air for 2.5 hours, then baked out a second time before Li deposition.  Air exposure would presumably not alter surface reconstructions due to the anneal, but difficult-to-remove atmospheric contaminants such as H$_2$O might return.

The intermediate air exposure  reduced $\Delta\rm{n}_{sat}$ back to $3\times10^{13}$ e$^-$/cm$^2$ [Fig.~5], not far above the unannealed value. This result suggests that limits on $\Delta\rm{n}_{sat}$ are primarily due to atmospheric gases absorbed on the graphene, rather than defects that are healed by high temperature annealing in UHV\cite{annealing.self.repair.Nano.Lett.2011, annealing.self.repair.Nat.Commun.2014, annealing.self.repair.ACS.Nano2015, annealing.self.repair.ACS.Nano2015.2, 2013IEEE.Conference}. However, the increase in $\Delta\rm{n}_{sat}$ from 700 to 900 K annealing temperature indicates these adsorbates are not fully removed even by anneals up to 700 K. This observation is difficult to reconcile with data from several groups indicating desorption temperatures for H$_2$O on graphene from 150 K to 400 K \cite{2012arXiv.water.desorption, annealing.water.desorption.J.Vac.Sci.Technol.B.2016, Smith:2014jh}.
It is well established that H$_2$O, O$_2$, H$_2$, and N$_2$ intercalate between monolayer graphene and its substrate (in many cases SiC) \cite{Water.intercalation.ACS.Appl.Mater.Interfaces2015, annealing.water.desorption.J.Vac.Sci.Technol.B.2016, Water.intercalation.J.Appl.Phys.2015, Water.intercalation.JACS.2012, Water.intercalation.nano.lett.2016, H.intercalation.J.Appl.Phys.2011, H.intercalation.PhysRevLett.2009, H.intercalation.NanoLett.2011, H.intercalation.J.Appl.Phys.2013,N2.intercalation.PhysRevB.2015, O2.water.intercalation.Carbon.2014, O2.intercalation.Carbon.2014}. On the other hand, it is not straightforward to identify a mechanism by which these intercalants would affect the doping efficiency of surface Li adatoms.

Although the data above have focused on $\Delta\rm{n}$, annealing also affected the pre-deposition carrier density $\rm{n}_{\rm{init}}$. Annealing above 500 K increased the  carrier density in bare SiC samples from $1-1.7\times10^{13}$ e$^-$/cm$^2$ to $2-2.7\times10^{13}$ e$^-$/cm$^2$, though exact values varied sample to sample.  After a first Li deposition step, annealing at 500 K brought $\rm{n}_{\rm{init}}$ close to its pre-Li value.  However, some indications were found that the desorption process for Li even after a 900 K anneal was not complete; this desorption will be the topic of a future study.

The data reported here motivate a more careful examination of the graphene-vacuum interface under UHV conditions, and how it evolves during annealing, via surface-sensitive techniques such as STM or LEED.  At the same time, they demonstrate that reliable adatom-graphene interactions can be achieved only after {\it in situ} high temperature cleaning procedures.  This observation may explain the difficulty many groups have faced in inducing superconductivity, spin-orbit interaction, or similar electronic modifications to graphene by adatom deposition, and points toward a straightforward, if experimentally challenging, solution.





\begin{acknowledgments}
The authors acknowledge A Damascelli, B Ludbrook, S Burke, G Levy, P Nigge, and A Macdonald for numerous discussions, as well as Ludbrook and J Renard for assistance in building the chamber.  A.K. thanks UBC for financial support through the Four Year Doctoral Fellowship.  Research supported by NSERC, CFI, and the SBQMI partnership with MPI.
\end{acknowledgments}

\end{document}